\documentclass{ws-ijmpa}
\usepackage[super,compress]{cite}
\usepackage{graphicx}
\usepackage{float}  
\begin{document}
\markboth{W. Lu}{Chiral symmetry breaking and fermion mass hierarchies}

%%%%%%%%%%%%%%%%%%%%% Publisher's Area please ignore %%%%%%%%%%%%%%%
%
\catchline{}{}{}{}{}
%
%%%%%%%%%%%%%%%%%%%%%%%%%%%%%%%%%%%%%%%%%%%%%%%%%%%%%%%%%%%%%%%%%%%%

\title{A Clifford Algebra Approach to Chiral Symmetry Breaking and Fermion Mass Hierarchies}

\author{Wei Lu}

\address{Forest Hills, NY 11375, USA\\
weiluphys@yahoo.com}

\maketitle

\begin{history}
\received{9 August 2017}
\revised{30 August 2017}
\end{history}

\begin{abstract}
We propose a Clifford algebra approach to chiral symmetry breaking and fermion mass hierarchies in the context of composite Higgs bosons. Standard model fermions are represented by algebraic spinors of six-dimensional binary Clifford algebra,  while ternary Clifford algebra-related flavor projection operators control allowable flavor-mixing interactions. There are three composite electroweak Higgs bosons resulted from top quark, tau neutrino, and tau lepton condensations. Each of the three condensations gives rise to masses of four different fermions. The fermion mass hierarchies within these three groups are determined by four-fermion condensations, which break two global chiral symmetries. The four-fermion condensations induce axion-like pseudo-Nambu-Goldstone bosons and can be dark matter candidates. In addition to the 125 GeV Higgs boson observed at the Large Hadron Collider, we anticipate detection of tau neutrino composite Higgs boson via the charm quark decay channel.

\end{abstract}

\ccode{PACS numbers:11.30.Rd, 11.30.Qc, 12.15.Ff, 12.60.Rc, 14.80.Va}

%11.30.Ly	Other internal and higher symmetries
%11.15.Ex	Spontaneous breaking of gauge symmetries
%11.30.Hv	Flavor symmetries
%11.30.Qc	Spontaneous and radiative symmetry breaking
%11.30.Rd	Chiral symmetries
%12.15.Ff	Quark and lepton masses and mixing
%12.60.Rc	Composite models
%14.60.Pq	Neutrino mass and mixing
%14.60.St	Non-standard-model neutrinos, right-handed neutrinos, etc
%14.80.Va	Axions and other Nambu-Goldstone bosons (Majorons, familons, etc.)

%\tableofcontents

\section{Introduction}	

Dimensionless ratios between parameters appearing in a physical theory can not be accidentally small. This naturalness principle is elegantly defined by 't Hooft\cite{THOO}: a quantity should be small only if the underlying theory becomes more symmetric as that quantity tends to zero. Weakly broken symmetry ensures that the smallness of a parameter is preserved against possible perturbative disturbances. The standard model Higgs sector is unnatural since even if one takes the massless Higgs boson limit, the symmetry of standard model is not enhanced. Perturbative quantum corrections tend to draw the smaller electroweak scale towards Planck scale. 

One way of addressing the naturalness problem is to replace the fundamental Higgs boson with a fermion-antifermion condensation, such as in technicolor\cite{TC1, TC2, TC3} and top condensation models\cite{TOP1, TOP2, TOP3, TOP4, TOP5, NEU1, NEU2, NEU3, NEU4, NEU5}. The Higgs sector is an effective description of the low energy physics represented by composite boson field. The condensation is induced via dynamical symmetry breaking mechanism, which is a profound concept in physics. It is introduced into relativistic quantum field theory by Nambu and Jona-Lasinio (NJL)\cite{NJL}, inspired by earlier Bardeen-Cooper-Schriefer (BCS) theory of superconductivity\cite{BCS}.

A challenge facing the composite Higgs model is to account for the vast range of fermion masses which span five orders of magnitude. The current paper is an effort towards explaining the fermion mass hierarchies in the context of composite electroweak Higgs bosons. We propose two global chiral symmetries, $U(1)_\alpha$ and $U(1)_\beta$, in addition to the local gauge symmetries. The chiral symmetries are dynamically broken by four-fermion condensations. In accordance with naturalness principle, the chiral symmetries play a pivotal role in establishing the relative magnitudes of four-fermion condensations, and consequently giving rise to fermion mass hierarchies.

Our approach is based on the framework of six-dimensional Clifford algebra $C\!\ell_{0,6}$ \cite{WL1, WL2}. Clifford algebra, also known as geometric algebra or space-time algebra (for the specific case of $C\!\ell_{1,3}$), is a powerful mathematical tool with various applications in physics\cite{HEST1,BAYL,TRAY,PAV,DORA,RODR}. %It was suggested by Hestenes\cite{HEST2} that the algebraic spinor of four-dimensional space-time algebra can represent electron and neutrino as two ideals. Electroweak gauge symmetries are readily accommodated by the algebraic spinor. A remarkable feature is the exact match of degree of freedom between an algebraic spinor ($2^4 = 16$) and that of an electron plus a neutrino ($8 + 8 = 16$). 
Including right-handed neutrinos, there are 16 Weyl fermions with $16\times2=32$ complex components ($64$ real components) within each of the three fermion families. One generation of fermions can be represented by an algebraic spinor, which is a linear combination of all $2^{6}=64$ basis elements of six-dimensional Clifford algebra $C\!\ell_{0,6}$. The $SU(3)_c \times SU(2)_{L} \times U(1)_{R} \times U(1)_{B-L}$ local gauge symmetries, which encompass standard model symmetries, are naturally embedded in the algebraic structure. Besides the binary Clifford algebra, ternary Clifford algebra\cite{NA1, NA2} is also leveraged in constructing flavor projection operators. They serve the purpose of determining allowable flavor-mixing interactions.

This paper is structured as follows: Section 2 introduces algebraic spinors and local gauge symmetries. In section 3, we study flavor-mixing interactions, chiral symmetry breaking by fermion condensations, and fermion mass hierarchies. In the last section we draw our conclusions.

\section{Clifford Algebra and Gauge Symmetries}

Standard model fermions (plus right-handed neutrinos) can be represented by algebraic spinors of six-dimensional Clifford algebra $C\!\ell_{0,6}$. The Lagrangian of the algebraic spinors accommodates local gauge symmetries $SU(3)_c \times SU(2)_{L} \times U(1)_{R} \times U(1)_{B-L}$. This section is a review of the algebraic spinor of $C\!\ell_{0,6}$ and the related symmetries. More details, including specifics about the mappings between conventional matrix representation and $C\!\ell_{0,6}$ formulation, can be found in Refs. \citen{WL1, WL2}.

\subsection{Fermions as algebraic spinors}

The six-dimensional Clifford algebra $C\!\ell_{0,6}$ is defined by vector basis $\{\Gamma_{j}; j= 1, 2, \ldots, 6\}$ satisfying
\begin{align}
&\Gamma_{j}\Gamma_{k}+\Gamma_{k}\Gamma_{j} = -2\delta_{jk}.
\end{align}
The four-dimensional Dirac space-time algebra $C\!\ell_{1,3}$ is a sub-algebra of $C\!\ell_{0,6}$ with basis
\begin{align}
&\gamma_{0} = \Gamma_{1}\Gamma_{2}\Gamma_{3}, \\
&\gamma_{1} = \Gamma_{4}, \\
&\gamma_{2} = \Gamma_{5}, \\
&\gamma_{3} = \Gamma_{6}.
\end{align}
Rather than being a vector, $\gamma_{0}$ here is a composite trivector, which departs from the other Clifford algebra-based approaches. Given the association of $\gamma_0$ with time dimension, ($\Gamma_1, \Gamma_2, \Gamma_3$) can be figuratively regarded as cube roots of time dimension. We will encounter ($\Gamma_1, \Gamma_2, \Gamma_3$) related trialities when we explore the three colors of quarks below and three bivectors of the left-handed weak gauge fields in the next subsection. That being said, as illustrated later in this paper, the three generations of fermions and three composite Higgs bosons are connected with a different sort of triality depicted by ternary Clifford algebra.

Color projection operators are given by 
\begin{align}
P_{rd} &= \frac{1}{4}(1+i\gamma_1\Gamma_1-i\gamma_2\Gamma_2-i\gamma_3\Gamma_3), \label{IDEM3}\\
P_{gr} &= \frac{1}{4}(1-i\gamma_1\Gamma_1+i\gamma_2\Gamma_2-i\gamma_3\Gamma_3), \label{IDEM4}\\
P_{bl} &= \frac{1}{4}(1-i\gamma_1\Gamma_1-i\gamma_2\Gamma_2+i\gamma_3\Gamma_3), \label{IDEM5}\\
P_l &= \frac{1}{4}(1+i\gamma_1\Gamma_1+i\gamma_2\Gamma_2+i\gamma_3\Gamma_3), \label{IDEM2}
\end{align}
where $i$ is the unit pseudoscalar
\begin{equation}
i =  \Gamma_{1}\Gamma_{2}\Gamma_{3}\Gamma_4\Gamma_5\Gamma_6=\gamma_0\gamma_{1}\gamma_{2}\gamma_{3},
\end{equation}
which squares to $-1$, anticommutes with Clifford-odd elements, and commutes with Clifford-even
elements. 

The lepton projection operator $P_l$ can be regarded as projection to the fourth color.
The quark projection operator $P_q$ is the sum of red, green, and blue projections
\begin{align}
&P_q = P_{rd}+P_{gr}+P_{bl}. \label{IDEM1}
\end{align} 
Additionally, we introduce another set of projection operators 
\begin{align}
&P_{\pm} = \frac{1}{2}(1\pm i{\Gamma_1}{\Gamma_2}), \label{IDEM6}
\end{align}
for the purpose of differentiating between weak isospin up-type and down-type fermions. 

One generation of fermions can be represented by an algebraic spinor, which is a linear combination of all $2^{6}=64$ basis elements of Clifford algebra $C\!\ell_{0,6}$. Due to the fermion nature, the linear combination coefficients are real Grassmann numbers. Note that the algebraic spinor is a $C\!\ell_{0,6}$-valued Grassmann-odd function of {\it four-dimensional} space-time, albeit $C\!\ell_{0,6}$ is six-dimensional. 

Spinors with left/right chirality correspond to Clifford-odd/even multivectors
\begin{align}
&\psi^a_{L} = \frac{1}{2}(\psi^a + i\psi^a i), \\
&\psi^a_{R} = \frac{1}{2}(\psi^a - i\psi^a i),
\end{align}
where three generations of spinors are denoted as $\psi^a$, with $a= 1,2,3$.
We identify projections of spinors 
\begin{equation}
\psi^a = (P_+ + P_-)(\psi^a_L + \psi^a_R)(P_q + P_l)
\end{equation}
%\vspace{10pt}
with quarks and leptons as shown in table~\ref{fermions}. 

\begin{table}[H]
\tbl{Three generations of fermions as projections of algebraic spinors. Quarks stand for sum of red, green, and blue colors. Individual colors of quarks can be obtained by applying color projection operators (\ref{IDEM3}), (\ref{IDEM4}), or (\ref{IDEM5}) to the quarks in the table. Going forward in this paper, quarks always denote the sum of three colors.}
{\begin{tabular}{|l|l|l|}
\hline
First Generation & Second Generation & Third Generation \\ 
\hline
&& \\
$u_L = P_+\psi^1_LP_q$ & $c_L = P_+\psi^2_LP_q$ & $t_L = P_+\psi^3_LP_q$ \\ 
&& \\
$d_L = P_-\psi^1_LP_q$ & $s_L = P_-\psi^2_LP_q$ & $b_L = P_-\psi^3_LP_q$ \\ 
&& \\
$\nu_{eL} = P_+\psi^1_LP_l$ & $\nu_{{\mu}L} = P_+\psi^2_LP_l$ & $\nu_{{\tau}L} = P_+\psi^3_LP_l$ \\ 
&& \\
$e_L = P_-\psi^1_LP_l$ & ${\mu}_L = P_-\psi^2_LP_l$ & ${\tau}_L = P_-\psi^3_LP_l$ \\
&& \\
\hline
&& \\
$u_R = P_-\psi^1_RP_q$ & $c_R = P_-\psi^2_RP_q$ & $t_R = P_-\psi^3_RP_q$ \\ 
&& \\
$d_R = P_+\psi^1_RP_q$ & $s_R = P_+\psi^2_RP_q$ & $b_R = P_+\psi^3_RP_q$ \\ 
&& \\
$\nu_{eR} = P_-\psi^1_RP_l$ & $\nu_{{\mu}R} = P_-\psi^2_RP_l$ & $\nu_{{\tau}R} = P_-\psi^3_RP_l$ \\ 
&& \\
$e_R = P_+\psi^1_RP_l$ & ${\mu}_R = P_+\psi^2_RP_l$ & ${\tau}_R = P_+\psi^3_RP_l$ \\
&& \\
\hline
\end{tabular}  \label{fermions}}
\end{table}

\subsection{Gauge fields and covariant derivatives}
The $SU(3)_c \times SU(2)_{L} \times U(1)_{R} \times U(1)_{B-L}$ gauge-covariant derivatives of fermion fields are
\begin{align}
&D_{L\mu}\psi^a_{L} = (\partial_{\mu} + W_{L\mu})\psi^a_{L} + \psi^a_{L} (W_{BL\mu} + G_\mu), \\
&D_{R\mu}\psi^a_{R} = (\partial_{\mu} + W_{R\mu})\psi^a_{R} + \psi^a_{R} (W_{BL\mu} + G_\mu),
\end{align}
where $\mu = 0, 1, 2, 3$. Here the gauge fields are defined to absorb gauge coupling constants.

The $SU(3)_c$ strong interaction $G_\mu$ is expressed as (summation convention for repeated indices is adopted in this paper)
\begin{align}
&G_{\mu} = G^k_{\mu}T_k,
\end{align}
where
\begin{equation}
\left(
\begin{array}{rl}
T_1, \ldots, T_8
\end{array}
\right)
= 
\left(
\begin{array}{rl}
&\frac{1}{4}(\gamma_1\Gamma_2 + \gamma_2\Gamma_1), 
\frac{1}{4}(\Gamma_1\Gamma_2 + \gamma_1\gamma_2), 
\frac{1}{4}(\Gamma_1\gamma_1 - \Gamma_2\gamma_2), \\
&\frac{1}{4}(\gamma_1\Gamma_3 + \gamma_3\Gamma_1),
\frac{1}{4}(\Gamma_1\Gamma_3 + \gamma_1\gamma_3),  \\
&\frac{1}{4}(\gamma_2\Gamma_3 + \gamma_3\Gamma_2),
\frac{1}{4}(\Gamma_2\Gamma_3 + \gamma_2\gamma_3),  \\
&\frac{1}{4\sqrt{3}}(\Gamma_1\gamma_1 + \Gamma_2\gamma_2 - 2\Gamma_3\gamma_3).
\end{array}
\right)   \label{SU3}
\end{equation}

The $SU(2)_{L}$ left-handed weak interaction $W_{L\mu}$, $U(1)_{R}$ right-handed weak interaction $W_{R\mu}$, and $U(1)_{B-L}$ interaction $W_{BL\mu}$ are of the form
\begin{align}
&W_{L\mu}= \frac{1}{2}(W^1_{L\mu}\Gamma_2\Gamma_3 + W^2_{L\mu}\Gamma_1\Gamma_3 + W^3_{L\mu}\Gamma_1\Gamma_2), \\
&W_{R\mu}= \frac{1}{2}W^3_{R\mu}\Gamma_1\Gamma_2, \\
&W_{BL\mu}= \frac{1}{2}W^J_{BL\mu}J, 
\end{align}
where 
\begin{align}
&J = \frac{1}{3}(\gamma_1\Gamma_1 + \gamma_2\Gamma_2 + \gamma_3\Gamma_3).
\end{align}
Thanks to the properties
\begin{align}
&P_{q}J = \frac{1}{3}P_{q}i, \label{BL1}\\
&P_lJ = -P_li, \label{BL2}
\end{align}
$W_{BL\mu}$ is equivalent to
\begin{align}
& W_{BL\mu} = \frac{1}{2}W^J_{BL\mu}(B-L)i, 
\end{align}
where $B$ and $L$ are baryon and lepton numbers, respectively. 

It can be verified that the product of lepton projector $P_l$ with any generator in color algebra (\ref{SU3}) is zero 
\begin{align}
&P_lT_k = 0.
\end{align}
As a result, leptons are $SU(3)_c$ singlets. They do not interact with gluons. 

After dynamic symmetry breaking of $SU(2)_{L} \times U(1)_{R} \times U(1)_{B-L}$, which will be discussed in later section, the remaining massless interactions are $G_\mu$ and electromagnetic field $A_\mu$. The $A_\mu$ part of gauge-covariant derivative is cast into the form
\begin{align}
&D_\mu\psi^a = (\partial_{\mu} + \frac{1}{2}A_\mu\Gamma_1\Gamma_2)\psi^a + \psi^a (\frac{1}{2}A_\mu J).
\end{align}
Electromagnetic field is non-chiral and makes no distinction between left- and right-handed spinors. Given (\ref{BL1}), (\ref{BL2}), and another property, 
\begin{align}
& \Gamma_1\Gamma_2P_{\pm} = {\mp} iP_{\pm},
\end{align}
the standard model electromagnetic charges for all individual fermions can thus be correctly derived\cite{WL1, WL2}.

\subsection{Gauge-invariant Lagrangian}

The gauge-invariant Lagrangian reads 
\begin{align}
\mathcal{L}_{World} =  &\mathcal{L}_{Fermion} + \mathcal{L}_{Yang-Mills} + \mathcal{L}_{Gravity} + \mathcal{L}_{Multi-Fermion}.
\end{align}
The fermion Lagrangian can be written as
\begin{equation}
\mathcal{L}_{Fermion} = \hat i\left\langle \bar{\psi}^a_L \gamma^{\mu} D_{L\mu} \psi^a_L
+ \bar{\psi}^a_R \gamma^{\mu} D_{R\mu} \psi^a_R \right\rangle, \label{ELEC}
\end{equation}
where %$\bar{\psi}^a_{L/R}$ denotes $(\psi^{a}_{L/R})^\dagger\gamma_0$, and 
$\gamma^{\mu} = \eta^{\mu\nu}\gamma_{\nu}$ ($\eta^{\mu\nu} = diag(1, -1, -1, -1)$), $\left\langle \ldots\right\rangle$ stands for Clifford-scalar part of enclosed expression, and $\bar{\psi}^a_{L/R}$ is defined as
\begin{equation}
\bar{\psi}^a_{L/R} = (\psi_{L/R}^a)^{\dagger}\gamma_0. 
\end{equation}
Hermitian conjugate $(\psi_{L/R}^a)^{\dagger}$ takes the form
\begin{equation}
(\psi_{L/R}^a)^{\dagger} = -i\tilde{\psi}^a_{L/R}i = \mp\tilde{\psi}^a_{L/R},
\end{equation}
where reversion of $\psi^a_{L/R}$, denoted $\tilde{\psi}^a_{L/R}$, reverses the
order in any product of Clifford vectors.

Note that $\hat i$ in the fermion Lagrangian is the mathematical imaginary number. It is different from Clifford algebra $C\!\ell_{0,6}$ pseudoscalar $i$. Imaginary number $\hat i$ commutes with all Clifford algebra elements. 

The local gauge symmetries can be extended to $SO(1,3)_{Lorentz} \times SU(3)_c \times SU(2)_{L} \times U(1)_{R} \times U(1)_{B-L}$ in the unified theory of gravity and Yang-Mills interactions\cite{WL2}. Gravity is treated as gauge theory of local Lorentz symmetry (see Refs. \citen{LORE, POIN, DESI, HEHL} for various gauge gravity theories). From an effective field theory point of view, an infinite number of terms allowed by symmetry requirements should be included in a generalized Lagrangian. The gravity and Yang-Mills Lagrangians are the first few order terms\cite{WL1} that are relevant in low-energy limit. The gravity gauge fields (vierbein/tetrad and spin connection fields) are best described by Clifford-valued one-forms or Clifforms\cite{WL1, WL2}. In a vacuum with zero cosmological constant, Minkowskian flat space-time is characterized by the nonzero vacuum expectation value (VEV) of vierbein/tetrad field, while VEV of spin connection is zero. The VEVs break the independent local Lorentz gauge symmetry and diffeomorphism invariance. The Lagrangian is left with a residual global Lorentz symmetry, corresponding to synchronized Clifford space and $x$ coordinate space global Lorentz rotations.

The gauge theory of gravity would also facilitate study of modified Einstein-Cartan gravity and its implications for cosmology.  The resultant modified Friedmannian cosmology (MFC)\cite{MODC} departs from the conventional Friedmannian cosmology for small Hubble parameter $H=\frac{\dot a(t)}{a(t)}=\frac{da(t)/dt}{a(t)}$, where $a(t)$ is the Robertson-Walker metric scale factor. A characteristic Hubble scale $h_0$ marks the boundary between the validity domains of Friedmannian cosmology and MFC. For large Hubble parameter $H \gg h_0$, Friedmannian cosmology is restored. In the opposite limit of small Hubble parameter (MFC regime: $H \lesssim h_0$), which includes the case of present epoch ($H_0 = \frac{\dot a(t_0)}{a(t_0)} \sim h_0$), Lorentz-violating effects would manifest themselves. One implication of MFC is that there may be no need to invoke dark matter to account for cosmological mass discrepancies. Another interesting observation is that MFC can accommodate late-time cosmic acceleration without cosmological constant.

We are not going to get into the details of gravity and Yang-Mills Lagrangians in this paper. We will instead focus on the investigation of multi-fermion interactions $\mathcal{L}_{Multi-Fermion}$ in later section. The multi-fermion interactions (and for that matter all the standard model gauge interactions) might be of gravitational origin. There are, among others, two approaches relating standard model with gravity. In the spin gauge theory of gravity\cite{PAV}, space-time itself can be replaced by Clifford space (C-space). A curved C-space provides a realization of Kaluza-Klein theory without the necessity of enlarging the dimensionality of space-time. The generalized spin connection in C-space has the properties of Yang-Mills gauge fields. It contains the ordinary spin connection related to gravity with torsion, and extra parts describing additional interactions, including those described by the antisymmetric Kalb-Ramond fields.

The inter-linkage between gravity field and standard model has also been studied in a geometrical five-dimensional approach\cite{CAPOZ2}, which deduces all the known interactions from an induced symmetry breaking of the non-unitary $GL(4)$-group of diffeomorphism. By a reduction procedure, the approach is capable of generating the masses of particles and their organization in families. The standard model is fully recovered by enlarging the gravitational sector, avoiding the Higgs boson and the hierarchy problem. The electroweak bosons are not gauge bosons in standard sense, whereas they can be derived from the gravitational degrees of freedom.

\subsection{The necessity of imaginary number}
%\subsection{Is Quantum Theory Real?}
The mathematical imaginary number $\hat i$ is ubiquitous in physics theories. The original objective of Clifford algebra (or geometric algebra) approach to physics is to abandon the imaginary number and replace it with certain even element in Clifford algebra. This initiative, pioneered by Hestenes\cite{HEST1}, has been fairly successful in a wide variety of physics domains, such as rotational symmetries, Dirac equation, gauge field theories, and quantum theory in the first quantization form. It's why we denote $C\!\ell_{0,6}$ pseudoscalar as $i$ in the first place. 

When it comes to quantum field theory (QFT) beyond tree approximation (with loop corrections), the jury is still out with regard to the status of imaginary number. As we have witnessed in the last subsection, imaginary number $\hat i$ appears in the fermion Lagrangian. Can we replace it with $C\!\ell_{0,6}$ pseudoscalar $i$? The answer is no. Given the Grassmann-odd nature of the spinor field, it turns out that an action in the form of
\begin{equation}
S_{Fermion} = \int {\left\langle i\bar{\psi}^a_L \gamma^{\mu} D_{L\mu} \psi^a_L
+ i\bar{\psi}^a_R \gamma^{\mu} D_{R\mu} \psi^a_R \right\rangle d^{4}x}
\end{equation}
can be proved to be equivalent to zero, after omission of surface integral terms. 

One might reckon that what really matters in the path integral formalism of QFT is
\begin{equation}
e^{\hat i \int { \mathcal{L} d^{4}x}},
\end{equation}
with additional source terms added in the Lagrangian. Since we know that
\begin{equation}
\hat i\mathcal{L}_{Fermion} = -\left\langle \bar{\psi}^a_L \gamma^{\mu} D_{L\mu} \psi^a_L
+ \bar{\psi}^a_R \gamma^{\mu} D_{R\mu} \psi^a_R \right\rangle
\end{equation}
is real, shall we claim that we can do away with imaginary number after all? The answer is still no. 

It's known that QFT propagators have poles. They are not properly defined without a prescription on integral in the vicinity of the poles. The beautiful Lorentz-invariant Feynman propagator hinges on the contour integral on the complex plane. Feynman's $\hat i \epsilon$ trick introduces the imaginary number through the back door. Equating the imaginary number with a Clifford algebra element in this context would seem rather unnatural\footnote{One can introduce a separate Clifford bivector $\hat i = \Gamma_7\Gamma_8$ to the original $C\!\ell_{0,6}$, so that it commutes with all $C\!\ell_{0,6}$ elements. However, this approach is not conducive to any additional physics insight.}. 

The interaction part of the Lagrangian is real though, be it Yang-Mills or multi-fermion interaction. Nevertheless, QFT loop integral would pick up an extra $\hat i$, via proper contour integral on the complex plane (or equivalently Wick rotation of time axis). Therefore, a self-energy loop diagram yields an imaginary correction to $\mathcal{L}_{Fermion}$. Unless we come up with some other innovative ways of performing integral around the propagator poles, we have to live with the imaginary number $\hat i$.

\section{Chiral Symmetries and Fermion Mass Hierarchies}
%\section{Multi-Fermion Interactions and Chiral Symmetry Breaking}

Dynamical symmetry breaking (DSB) is introduced into relativistic quantum field theory by Nambu and Jona-Lasinio\cite{NJL}. The NJL model is based on a four-fermion interaction, which is strong enough to induce fermion-antifermion condensation via DSB mechanism. Multi-fermion interactions are not renormalizable in the conventional sense. They can be regarded as effective representations of underlying renormalizable theory.

We subscribe to the general notion that multi-fermion interactions $\mathcal{L}_{Multi-Fermion}$ are instrumental in driving DSB and giving rise to four-fermion as well as two-fermion condensations. The bosonic sector is just an effective Ginzburg-Landau-type description of the low energy physics represented by composite boson fields. 

Two global chiral symmetries are propounded in this section. They play the crucial role of dictating the relative magnitudes of four-fermion condensations, and consequently shaping fermion mass hierarchies.

\subsection{Flavor projection operators}
For the purpose of investigating allowable flavor-mixing multi-fermion interactions $\mathcal{L}_{Multi-Fermion}$, we resort to another kind of Clifford algebra involving ternary communication relationships\cite{NA1, NA2} rather than the usual binary ones. Let's consider a ternary Clifford algebra with a single vector $\zeta$ satisfying
\begin{align}
&\zeta^3 = 1,
\end{align}
with $\zeta$ commuting with $C\!\ell_{0,6}$. We introduce three projection operators which involve both binary and ternary Clifford algebra elements
\begin{align}
\zeta^{0} &= \frac{1}{3}(1+\zeta+\zeta^2), \\
\zeta^{+} &= \frac{1}{3}(1+e^{\frac{2\pi}{3}i}\zeta+e^{-\frac{2\pi}{3}i}\zeta^2),  \\
\zeta^{-} &= \frac{1}{3}(1+e^{-\frac{2\pi}{3}i}\zeta+e^{\frac{2\pi}{3}i}\zeta^2).
\end{align}
Flavor projection operators for the three families of fermions are defined by\cite{WL2}
\begin{align}
P^{1} &= P_q\zeta^{-} + P_l\zeta^{0},  \label{IDEM7}\\
P^{2} &= P_q\zeta^{+} + P_l\zeta^{-},  \label{IDEM8}\\
P^{3} &= P_q\zeta^{0} + P_l\zeta^{+},  \label{IDEM9}
\end{align}
where $P_q$ and $P_l$ are quark and lepton projection operators, respectively. 

Note that ($\zeta^{0}$, $\zeta^{\pm}$) operators are assigned to quarks and leptons in disparate patterns. As demonstrated in later subsections, this particular layout is structured to facilitate mixing between first and second generation quarks as well as second and third generation leptons. The mixing stems from properties
\begin{align}
&\zeta^{0}A = A\zeta^{0}, \\
&\zeta^{+}A = A\zeta^{-}, \\
&\zeta^{-}A = A\zeta^{+},
\end{align}
for any $C\!\ell_{0,6}$-odd element $A$, since pseudoscalar $i$ in the definition of ($\zeta^{0}$, $\zeta^{\pm}$) operators anticommutes with $C\!\ell_{0,6}$-odd $A$. On the other hand, ($\zeta^{0}$, $\zeta^{\pm}$) operators commute with $C\!\ell_{0,6}$-even elements.

We know that flavor mixing is observed between all generations, giving rise to distinct configurations of CKM and PMNS matrices. The above flavor projection assignment captures the most significant mixing effects. To allow for further mixing possibilities, we can potentially adopt other flavor projection assignments. The rule of thumb is that only one ansatz should be used for a given Lagrangian term, while separate terms are permitted to adopt different ansatzes. For the current paper, we will focus on the schema defined in (\ref{IDEM7}, \ref{IDEM8}, \ref{IDEM9}). The interaction terms subjected to other flavor projection assignments are presumably associated with suppressed coupling constants.  We leave the study of these subdued interactions to future research. 

If we take a step back and think twice about it, we would realize that the traditional way of generation assignment is rather arbitrary. For instance, there is no compelling rational for categorizing ($u$, $d$) and ($\nu_e$, $e$) into the same generation. The assignment is purely out of convenience, since ($u$, $d$, $e$) are the lightest bunch and constitute the bulk of building blocks of the visible universe. One might argue that the chiral anomaly cancellation condition relates quarks to leptons within a generation. Nonetheless, quantum anomaly cancellation requires equal numbers of quark and lepton generations. It does not tie a specific quark generation to a given lepton generation.

%It makes more sense to call a spade a spade by adopting the following naming convention,  
If one adopts following generation assignment convention,
\begin{align}
Generation \quad \zeta^{0} &: \quad t, b, \nu_e, e,  \label{Z0}\\
Generation \quad \zeta^{+} &: \quad c, s, \nu_{\tau}, \tau,  \label{ZP}\\
Generation \quad \zeta^{-} &: \quad u, d, \nu_{\mu}, \mu, \label{ZN}
\end{align}
the flavor projection operators are thus simply ($\zeta^{0}$, $\zeta^{\pm}$),  without expressly referencing quark and lepton projection operators. 

%It's not merely about playing nomenclature music chairs. 
It's not merely about changing the nomenclature. For example, we know that the algebraic spinor of $C\!\ell_{0,6}$ can actually accommodate $SU(4)$ gauge symmetry\cite{WL1, WL2}, which encompasses $SU(3)_c \times U(1)_{B-L}$. If nature allows for such gauge interaction, the coset $SU(4)/(U(3)_c \times U(1)_{B-L})$ related gauge fields (which are massive and Clifford-even) would transform down/up quarks into muons/muon neutrinos of the same $\zeta^{-}$ generation, instead of into electrons/electron neutrinos of $\zeta^{0}$ generation as usually assumed for proton decay. Because of the heavy muon mass, the tree-level amplitude for proton decay into meson and positive muon would therefore be suppressed.

%For the rest of the paper, proton decay is of no concern.  We are going to stay with the traditional generation naming convention.

Going forward, we will strictly follow the flavor projection regime which stipulates that fermions and flavor projection operators should always stick together in pairs as
\begin{align}
&\psi^1P^{1}, \psi^2P^{2},  \psi^3P^{3},  \\
&P^{1}\bar{\psi}^1, P^{2}\bar{\psi}^2, P^{3}\bar{\psi}^3.
\end{align}

Let's test the flavor projection rule with a generalized fermion kinetic Lagrangian
\begin{equation}
\mathcal{L}_{Fermion} = \hat i\left\langle P^{a}\bar{\psi}^a_L \gamma^{\mu} D_{L\mu} \psi^b_LP^{b}
+ P^{a}\bar{\psi}^a_R \gamma^{\mu} D_{R\mu} \psi^b_RP^{b} \right\rangle.
\end{equation}
Given the orthogonal properties 
\begin{align}
& \zeta^{0}\zeta^{-} = \zeta^{0}\zeta^{+} = \zeta^{-}\zeta^{+} = 0, 
\end{align}
one can easily verify that there is no flavor-mixing cross term in the Lagrangian. Flavor mixing is permitted for multi-fermion interactions, which will be studied in later subsections. 

For sake of simplifying notations, from now on we will not explicitly write down flavor projection operators. Each fermion field should implicitly assume an accompanying flavor projection operator.

%It should be noted that there is another Clifford algebra based approach called spin-charge-family theory\cite{BORS}. It predicts a fourth family, coupled to the observed three families. It also predicts a second group of four families, with the lowest of these four families explaining the origin of dark matter. 

\subsection{Right-handed-only four-fermion interactions}
The right-handed-only interactions are responsible for dynamically generating neutrino Majorana masses. The  $SU(3)_c \times SU(2)_{L} \times U(1)_{R} \times U(1)_{B-L}$ gauge-invariant interaction Lagrangian contains
\begin{align}
\mathcal{L}_{Majorana} =
&G_{11}\left\langle \bar{\nu}_{eR}\Gamma_2\Gamma_3\nu_{eR}\bar{\nu}_{eR}\Gamma_2\Gamma_3\nu_{eR} \right\rangle \label{NOTMIX}\\
+ &G_{23}\left\langle \bar{\nu}_{{\mu}R}\Gamma_2\Gamma_3\nu_{{\tau}R}\bar{\nu}_{{\tau}R}\Gamma_2\Gamma_3\nu_{{\mu}R} \right\rangle + h.c., \label{MIX}
\end{align}
where $G_{11}$ and $G_{23}$ are coupling constants. The bivector $\Gamma_2\Gamma_3$ can be replaced by arbitrary combination of $\Gamma_2\Gamma_3$ and $\Gamma_1\Gamma_3$. But it does not change the overall picture.

As stated earlier, flavor projection operators are implicitly attached to fermions. The permissible flavor-mixing patterns of interactions are controlled by the properties of flavor projection operators and weak isospin $P_{\pm}$ projection operators (note that $P_{-}\gamma_{0}\Gamma_2\Gamma_3 = \gamma_{0}\Gamma_2\Gamma_3P_{-}$). Other right-handed-only four-fermion interactions are also allowable. Some of them are examined in Ref. \citen{WL2}. Since they don't contribute to Majorana type two-fermion condensations, these additional terms are not enumerated here.

\subsection{Right-left-mixing four-fermion interactions}
Right-left-mixing interactions are employed by top condensation model\cite{TOP1, TOP2, TOP3, TOP4,TOP5} for dynamically breaking electroweak symmetry. The simplest version of top condensation model assumes top quark-antiquark condensation only. For the purpose of saturating electroweak scale, the scenario has been extended, among others\cite{TOP5}, to neutrino condensations\cite{NEU1, NEU2, NEU3, NEU4, NEU5} as well. 

Based on the flavor projection operator properties, three or more condensations are required to generate Dirac masses for all three generations of fermions.  We adopt the minimalist approach by allowing for top quark, tau neutrino, and tau lepton condensations only. Our premise is that any other fermion condensations are either nonexistent or negligible. As shown in later subsections, this unique ensemble of condensations is consistent with fermion mass hierarchies controlled by two chiral symmetries. 

The $SU(3)_c \times SU(2)_{L} \times U(1)_{R} \times U(1)_{B-L}$ gauge-invariant right-left-mixing interactions pertaining to these three condensations are
\begin{align}
\mathcal{L}_{t} &=
\frac{1}{4}g_{t}\left\langle \bar{q}^3_{L}\gamma^{\mu}{q}^3_{L}\bar{t}_{R}\gamma_{\mu}{t}_{R} \right\rangle \label{tt} \\
&+ \frac{1}{4}g_{t{\nu_e}}\left\langle \bar{l}^1_{L}\gamma^{\mu}{q}^3_{L}\bar{t}_{R}\gamma_{\mu}{\nu}_{eR} \right\rangle \label{tnu} + h.c. \\
&- g_{tb}\left\langle \bar{q}^3_{L}{q}^3_{L}\bar{t}_{R}{b}_{R} \right\rangle \label{tb} + h.c. \\
&- g_{te}\left\langle \bar{l}^1_{L}{q}^3_{L}\bar{t}_{R}{e}_{R} \right\rangle \label{te} + h.c., \\
\mathcal{L}_{\nu_{\tau}} &=
\frac{1}{4}g_{\nu_{\tau}}\left\langle \bar{l}^3_{L}\gamma^{\mu}{l}^3_{L}\bar{\nu}_{\tau R}\gamma_{\mu}{\nu_{\tau}}_{R} \right\rangle  \\
&+ \frac{1}{4}g_{\nu_{\tau}{c}}\left\langle \bar{q}^2_{L}\gamma^{\mu}{l}^3_{L}\bar{\nu}_{\tau R}\gamma_{\mu}{c}_{R} \right\rangle + h.c.  \\
&- g_{\nu_{\tau}\mu}\left\langle \bar{l}^2_{L}{l}^3_{L}\bar{\nu}_{\tau R}{\mu}_{R} \right\rangle + h.c.  \label{MIX23}\\
&- g_{\nu_{\tau}d}\left\langle \bar{q}^1_{L}{l}^3_{L}\bar{\nu}_{\tau R}{d}_{R} \right\rangle + h.c., \\
\mathcal{L}_{{\tau}} &=
\frac{1}{4}g_{{\tau}}\left\langle \bar{l}^3_{L}\gamma^{\mu}{l}^3_{L}\bar{{\tau}}_{R}\gamma_{\mu}{{\tau}}_{R} \right\rangle  \\
&+ \frac{1}{4}g_{{\tau}{s}}\left\langle \bar{q}^2_{L}\gamma^{\mu}{l}^3_{L}\bar{{\tau}}_{R}\gamma_{\mu}{s}_{R} \right\rangle + h.c.  \\
&- g_{{\tau}{\nu_\mu}}\left\langle \bar{l}^2_{L}{l}^3_{L}\bar{{\tau}}_{R}{\nu_\mu}_{R} \right\rangle + h.c.  \label{MIX32}\\
&- g_{{\tau}u}\left\langle \bar{q}^1_{L}{l}^3_{L}\bar{{\tau}}_{R}{u}_{R} \right\rangle  + h.c., 
\end{align}
where $g_{\ldots}$ are coupling constants. The left-handed doublets $q^a_{L}$ and $l^a_{L}$ are understood as the sum of weak isospin up-type and down-type fermions. For example, $q^1_{L}$ and $l^1_{L}$ denote
\begin{align}
&q^1_{L} = u_L + d_L, \\
&l^1_{L} = \nu_{eL} + e_L.
\end{align}

The permissible flavor-mixing patterns of these four-fermion terms are dictated by the properties of flavor projection operators and weak isospin $P_{\pm}$ projection operators. Given $P_{\pm}\gamma_{0}\gamma_{\mu} = \gamma_{0}\gamma_{\mu}P_{\pm}$, we know that $\bar{t}_{R}\gamma_{\mu}{t}_{R} = t^{\dagger}_{R}\gamma_0\gamma_{\mu}{t}_{R} = (P_{-}t_{R})^{\dagger}\gamma_0\gamma_{\mu}{P_{-}t}_{R} = t_{R}^{\dagger}P_{-}\gamma_0\gamma_{\mu}{P_{-}t}_{R}$ is non-zero.  Given $P_{\pm}\gamma_{0} = \gamma_{0}P_{\mp}$, we know that $\bar{t}_{R}{b}_{R} = t^{\dagger}_{R}\gamma_0{b}_{R} = (P_{-}t_{R})^{\dagger}\gamma_0{P_{+}b}_{R} = t_{R}^{\dagger}P_{-}\gamma_0{P_{+}b}_{R}$ is non-zero. Similar logic applies to other fermion combinations. 

In the $\zeta$ generation parlance, top cohort $\mathcal{L}_{t}$ corresponds to $\zeta^{0}$ generation, while tau neutrino and tau lepton cohorts $\mathcal{L}_{\nu_{\tau}}$ and $\mathcal{L}_{{\tau}}$ represent mixture of $\zeta^{\pm}$ generations. 

The vector interactions contain pairs $\gamma^{\mu}$/$\gamma_{\mu}$, while the others are scalar interactions. One can potentially add bivector interactions with pairs $\gamma^{\mu}\gamma^{\nu}$/$\gamma_{\mu}\gamma_{\nu}$, though they more or less behave in a similar manner as the scalar counterparts. As such, we will not write down bivector interactions separately. 

Other right-left-mixing four-fermion interactions are also allowed. For instance, the following flavor-changing charged interactions
\begin{align}
\mathcal{L}_{{Quark}-{Mixing}} &=g_{us}\left\langle \bar{q}^2_{L}{q}^1_{L}\bar{{u}}_{R}{s}_{R} \right\rangle + h.c. \\
&+ g_{cd}\left\langle \bar{q}^1_{L}{q}^2_{L}\bar{{c}}_{R}{d}_{R} \right\rangle + h.c.
\end{align}
mix first and second generation quarks. They flip isospin up-type quarks to down-type quarks, and {\it vice versa}.  They are not directly driving the three electroweak condensations in question. %We will not discuss these kinds of interactions further in this paper.
On the other hand, the lepton counterparts of flavor-changing charged interactions (\ref{MIX23}, \ref{MIX32})
mix second and third generation leptons. However, this flavor-mixing effect is overshadowed by the flavor-mixing originated from the right-handed-only four-neutrino interactions (\ref{MIX}) and the resultant flavor-mixing Majorana masses demonstrated in later subsection. This is the reason underlying the large mixing angles of PMNS matrix compared with CKM matrix (see Refs. \citen{TRI1, TRI2} for different attempts of explaining the PMNS matrix pattern).

Note that flavor-changing neutral interactions, such as
\begin{align}
\left\langle \bar{q}^2_{L}\gamma^{\mu}{q}^1_{L}\bar{{d}}_{R}\gamma_{\mu}{s}_{R} \right\rangle,
\end{align}
are identically zero, since for Clifford-even $\bar{{d}}_{R}\gamma_{\mu}{s}_{R}$ we have 
\begin{align}
P^1\bar{{d}}_{R}\gamma_{\mu}{s}_{R}P^2  = \zeta^-\bar{{d}}_{R}\gamma_{\mu}{s}_{R}\zeta^+ = \bar{{d}}_{R}\gamma_{\mu}{s}_{R}\zeta^-\zeta^+ = 0.
\end{align}

\subsection{Four-fermion condensations and chiral symmetry breaking}
As mentioned earlier, four-fermion interactions are invariant under $SU(3)_c \times SU(2)_{L} \times U(1)_{R} \times U(1)_{B-L}$ gauge transformations. These gauge symmetries are local and related to gauge interactions. In this subsection, we investigate two additional global chiral symmetries $U(1)_\alpha$ and $U(1)_\beta$ associated with right-handed fermions,
\begin{align}
U(1)_\alpha: \quad&\psi^a_R \quad\Rightarrow\quad \psi^a_Re^{\alpha i}, \\
U(1)_\beta: \quad&P_{-}\psi^a_RP_{q} \quad\Rightarrow\quad P_{-}\psi^a_RP_{q}e^{\beta i}, \\
 &P_{+}\psi^a_RP_{q} \quad\Rightarrow\quad P_{+}\psi^a_RP_{q}e^{-\beta i}, \\
 &P_{-}\psi^a_RP_{l} \quad\Rightarrow\quad P_{-}\psi^a_RP_{l}e^{-\beta i}, \\
 &P_{+}\psi^a_RP_{l} \quad\Rightarrow\quad P_{+}\psi^a_RP_{l}e^{\beta i}.
\end{align}
The $\alpha$-type chiral transformation rotates all right-handed fermions by the same phase $e^{\alpha i}$. The $\beta$-type chiral transformation rotates weak isospin up-type quarks ($u_R$, $c_R$, $t_R$) and down-type leptons ($e_R$, $\mu_R$, $\tau_R$) by $e^{\beta i}$, while it rotates down-type quarks ($d_R$, $s_R$, $b_R$) and up-type leptons ($\nu_{eR}$, $\nu_{{\mu}R}$, $\nu_{{\tau}R}$) by $e^{-\beta i}$. One can replace $U(1)_\alpha$ with $U(1)_A$:
\begin{align}
&\psi^a_R \Rightarrow \psi^a_Re^{\alpha i} = e^{\alpha i}\psi^a_R, \\
&\psi^a_L \Rightarrow \psi^a_Le^{-\alpha i} = e^{\alpha i}\psi^a_L,
\end{align}
without qualitatively affecting the following discussion.

Note that the definitions of chiral symmetries are not explicitly generation dependent, differing from earlier efforts of linking generation-dependent chiral symmetry\cite{FROG} with fermion mass hierarchy.

The fermion kinetic Lagrangian and right-handed-only four-fermion interactions are invariant under $U(1)_\alpha$ and $U(1)_\beta$ transformations. However, some of the right-left-mixing four-fermion interactions do not observe the chiral symmetries. The symmetry violation pattern is summarized in table~\ref{symmetries}. The top row respects both $U(1)_\alpha$ and $U(1)_\beta$ symmetries, while the bottom row violates both symmetries. On the Lagrangian level, $U(1)_\alpha$ invariance is controlled by whether or not two right-handed fermions are both isospin up-type or both down-type (equivalently, vector vs scalar interactions), while $U(1)_\beta$ invariance is determined by whether or not two right-handed fermions are both quarks or both leptons.

\begin{table}[H]
\tbl{The $U(1)_\alpha$ and $U(1)_\beta$ symmetry violation pattern grouped by top quark, tau neutrino, and tau lepton cohorts. Four-fermion interactions are represented by coupling constants along with the extra phases originated from the chiral transformations of right-handed fermions.}
{\begin{tabular}{|l|l|l|l|l|}
\hline
$t$ Cohort & $\nu_{\tau}$ Cohort & $\tau$ Cohort & $U(1)_\alpha$ & $U(1)_\beta$ \\ 
\hline
&&&& \\
$g_{t}$ & $g_{\nu_{\tau}}$ & $g_{{\tau}}$ & $\surd$ & $\surd$ \\ 
&&&& \\
$g_{t{\nu_e}}e^{- 2\beta i}$ & $g_{\nu_{\tau}{c}}e^{2\beta i}$ & $g_{{\tau}{s}}e^{- 2\beta i}$ & $\surd$ & $\times$ \\ 
&&&& \\
$g_{tb}e^{2\alpha i}$ & $g_{\nu_{\tau}\mu}e^{2\alpha i}$ & $g_{{\tau}{\nu_\mu}}e^{2\alpha i}$ & $\times$ & $\surd$ \\ 
&&&& \\
$g_{te}e^{2\alpha i + 2\beta i}$ & $g_{\nu_{\tau}d}e^{2\alpha i - 2\beta i}$ & $g_{{\tau}u}e^{2\alpha i + 2\beta i}$ & $\times$ & $\times$ \\
&&&& \\
\hline
\end{tabular} \label{symmetries}}
\end{table}

The four-fermion interactions can be rendered $U(1)_\alpha$ and $U(1)_\beta$ invariant, if we promote the coupling constants in the second, third, and fourth rows to composite boson fields. These composite boson fields are endowed with proper chiral charges to net out the extra phases in table~\ref{symmetries}. For example, the $g_{tb}$-related four-fermion interaction shall be
\begin{align}
\mathcal{L}_{tb} &= \Lambda^{-3}\left\langle \Phi_{tb} \bar{q}^3_{L}{q}^3_{L}\bar{t}_{R}{b}_{R} \right\rangle + h.c.,
\end{align}
where $\Lambda$ is a cutoff scale, and $\Phi_{tb}$ is valued in the Clifford space spanned by 
\begin{align}
&\{P_q\zeta^0, iP_q\zeta^0\}. 
\end{align}
The composite boson field $\Phi_{tb}$ is a $SU(3)_c \times SU(2)_{L} \times U(1)_{R} \times U(1)_{B-L}$ singlet. In terms of chiral symmetries, $\Phi_{tb}$ transforms as
\begin{align}
&\Phi_{tb} \Rightarrow \Phi_{tb}e^{-2\alpha i}.
\end{align}
As a result, the Lagrangian $\mathcal{L}_{tb}$ is chiral symmetry invariant. 

The chiral symmetry-related composite boson fields are effective representations of four-fermion condensations. The notion of four-fermion condensation has been proposed in Ref. \citen{WL2}. Four-fermion condensations produce various pseudo-Nambu-Goldstone bosons and can be dark matter candidates, since they are standard model singlets and don't directly interact with gauge fields. 

By virtue of DSB mechanism, the four-fermion condensations are induced by the underlying eight-fermion interactions which honor all $SU(3)_c \times SU(2)_{L} \times U(1)_{R} \times U(1)_{B-L} \times U(1)_\alpha \times U(1)_\beta$ symmetries. For instance, a $\Phi_{tb}/g_{tb}$-related eight-fermion interaction takes the form
\begin{align}
\mathbb{G}_{tb}\left\langle (\bar{b}_{R}{t}_{R}\bar{q}^3_{L}{q}^3_{L})(\bar{q}^3_{L}{q}^3_{L}\bar{t}_{R}{b}_{R}) \right\rangle,
\end{align}
where $\mathbb{G}_{tb}$ is the eight-fermion coupling constant. The four-fermion coupling constant $g_{tb}$ is effectively the magnitude of four-fermion condensation of 
\begin{align}
\Lambda^{-3}\Phi_{tb} &\quad\sim\quad  \mathbb{G}_{tb}\bar{b}_{R}{t}_{R}\bar{q}^3_{L}{q}^3_{L}\quad\rightarrow\quad g_{tb}P_q\zeta^0, 
\end{align}
which breaks the $U(1)_\alpha$/$U(1)_A$ symmetry, while leaving the $SU(3)_c \times SU(2)_{L} \times U(1)_{R} \times U(1)_{B-L}$ gauge symmetries intact. Note that due to the explicit chiral symmetry breaking originated from quantum anomaly and instanton effects, the pseudo-Nambu-Goldstone boson associated with the $U(1)_\alpha$/$U(1)_A$ symmetry breaking acquires a mass in a similar fashion as the axion\cite{AXN1, AXN2, AXN3}. Given that the four-fermion condensations are local gauge (especially electroweak) singlets, they are more in line with the invisible axion models\cite{INV1, INV2, INV3, INV4}. It is worthwhile to study further the four-fermion condensations as a possible solution to strong CP problem. 

The magnitudes of $U(1)_\alpha$ symmetry breaking four-fermion condensations are presumably smaller than the $U(1)_\beta$ counterparts, thus making $U(1)_\alpha$ the primary agent and $U(1)_\beta$ the secondary agent in establishing the hierarchies of four-fermion coupling constants.

We believe that the constants $g_{t}$, $g_{\nu_{\tau}}$, $g_{{\tau}}$, $G_{11}$, and $G_{23}$ are of the same order, since these couplings observe both chiral symmetries. In light of the symmetry breaking pattern of table~\ref{symmetries}, the coupling constants within the same column can be progressively smaller from the top row to the bottom row, in accordance with 't Hooft's naturalness principle. As stated above, we make the assumption that $U(1)_\alpha$ is primary and $U(1)_\beta$ is secondary. As a result, symmetry breaking of $U(1)_\alpha$ matters more, determining the order of the second and third row. It is supported by the fact that charm quark is heavier than muon as shown in next subsection.

\subsection{Two-fermion condensations and gauge symmetry breaking}

Two-fermion condensations are resulted from the four-fermion interactions $\mathcal{L}_{Majorana}$, $\mathcal{L}_{t}$, 
$\mathcal{L}_{\nu_{\tau}}$, and $\mathcal{L}_{{\tau}}$ studied in prior subsections. There are five composite Higgs bosons corresponding to two Majorana condensations,
\begin{align}
H_{11} &\quad\sim\quad  \Lambda^{-2}\hat i i\bar{\nu}_{eR}\Gamma_2\Gamma_3\nu_{eR} \quad\rightarrow\quad \frac{1}{2}\upsilon_{11}P_l\gamma_0\zeta^0, \\
H_{23} &\quad\sim\quad  \Lambda^{-2}\hat i i\bar{\nu}_{{\mu}R}\Gamma_2\Gamma_3\nu_{{\tau}R} \quad\rightarrow\quad \frac{1}{2}\upsilon_{23}P_l\zeta^-\gamma_0\zeta^+,
\end{align}
and three electroweak condensations,
\begin{align}
h_{t}  &\quad\sim\quad  \Lambda^{-2}\hat i{q}^3_{L}i\bar{t}_{R} \quad\rightarrow\quad \upsilon_{t}P_+\zeta^0, \\
h_{\nu_{\tau}}  &\quad\sim\quad  \Lambda^{-2}\hat i{l}^3_{L}i\bar{\nu}_{\tau R} \quad\rightarrow\quad \upsilon_{\nu_{\tau}}P_+\zeta^-, \\
h_{{\tau}}  &\quad\sim\quad  \Lambda^{-2}\hat i{l}^3_{L}i\bar{{\tau}}_{R} \quad\rightarrow\quad \upsilon_{{\tau}}P_-\zeta^-,
\end{align}
where $\upsilon_{11}$, $\upsilon_{23}$, $\upsilon_{t}$, $\upsilon_{\nu_{\tau}}$, and $\upsilon_{{\tau}}$ are condensation magnitudes, and $\Lambda$ is the cutoff scale.

The collective modes of composite Higgs fields can be determined as the poles of bosonic channels of the four-fermion interactions by summing to infinite order chains of bubble perturbation diagrams. The leading order calculation goes by different names such as random-phase approximation, Bethe-Salpeter T-matrix equation, or 1/N expansion.

Majorana Higgs bosons $H_{11}$ and $H_{23}$ correspond to right-handed neutrino condensations. The Majorana condensations break the symmetries from $SU(3)_c \times SU(2)_{L} \times U(1)_{R} \times U(1)_{B-L}$ to the standard model symmetries $SU(3)_c \times SU(2)_{L} \times U(1)_Y$, where $U(1)_Y$ is the hypercharge gauge symmetry. Gauge field $Z'$ acquires a mass as a consequence\cite{WL2}. 

Electroweak Higgs bosons $h_{t}$ , $h_{\nu_{\tau}}$, and $h_{{\tau}}$ represent top quark, tau neutrino, and tau lepton condensations, respectively. The electroweak condensations break the symmetries from $SU(3)_c \times SU(2)_{L} \times U(1)_{R} \times U(1)_{B-L}$ to $SU(3)_c \times U(1)_{LR} \times U(1)_{B-L}$, where $U(1)_{LR}$ corresponds to the synchronization of right-handed weak gauge symmetry and third component of left-handed weak gauge symmetry. As a result, gauge fields $W^{\pm}$ and $Z$ gain masses. 

Collectively, these five composite Higgs bosons break the symmetries from $SU(3)_c \times SU(2)_{L} \times U(1)_{R} \times U(1)_{B-L}$ to $SU(3)_c \times U(1)_{em}$, where $U(1)_{em}$ is the electromagnetic gauge symmetry.

With fermion pairs in the four-fermion Lagrangians approximated by condensation values (e.g., via self-consistent Hartree-Fock gap equations), the resultant fermion mass terms are
\begin{align}
\Lambda^{-2}\mathcal{L}_{Mass} &= G_{11}\upsilon_{11}\hat i\left\langle i\gamma_0\bar{\nu}_{eR}\Gamma_2\Gamma_3\nu_{eR}\right\rangle \\
			&+ G_{23}\upsilon_{23}\hat i\left\langle i\gamma_0\bar{\nu}_{{\mu}R}\Gamma_2\Gamma_3\nu_{{\tau}R}\right\rangle + G_{23}\upsilon_{23}\hat i\left\langle i\gamma_0\bar{\nu}_{{\tau}R}\Gamma_2\Gamma_3\nu_{{\mu}R}\right\rangle \label{MIXX} \\
			&+ g_{t}\upsilon_{t}\hat i\left\langle i\bar{t}t\right\rangle
					+ g_{t\nu_e}\upsilon_{t}\hat i\left\langle i\bar{\nu}_e\nu_e \right\rangle
					+ g_{tb}\upsilon_{t}\hat i\left\langle i\bar{b}b\right\rangle
					+ g_{te}\upsilon_{t}\hat i\left\langle i\bar{e}e\right\rangle \\
			&+ g_{\nu_{\tau}}\upsilon_{\nu_{\tau}}\hat i\left\langle i\bar{\nu}_{\tau}\nu_{\tau}\right\rangle
					+ g_{\nu_{\tau}c}\upsilon_{\nu_{\tau}}\hat i\left\langle i\bar{c}c \right\rangle
					+ g_{\nu_{\tau}\mu}\upsilon_{\nu_{\tau}}\hat i\left\langle i\bar{\mu}\mu \right\rangle
					+ g_{\nu_{\tau}d}\upsilon_{\nu_{\tau}}\hat i\left\langle i\bar{d}d\right\rangle \\
			&+ g_{{\tau}}\upsilon_{{\tau}}\hat i\left\langle i\bar{{\tau}}{\tau}\right\rangle
					+ g_{{\tau}s}\upsilon_{{\tau}}\hat i\left\langle i\bar{s}s \right\rangle
					+ g_{{\tau}\nu_{\mu}}\upsilon_{{\tau}}\hat i\left\langle i\bar{\nu}_{\mu}\nu_{\mu} \right\rangle
					+ g_{{\tau}u}\upsilon_{{\tau}}\hat i\left\langle i\bar{u}u\right\rangle.
\end{align}
The first two lines are Majorana mass terms of neutrinos. The rest are Dirac mass terms. Note that imaginary number $\hat i$ in the mass terms comes from fermion condensations which pick up an extra $\hat i$ via self-energy contour integral on the complex plane (or equivalently Wick rotation of time axis). The conventional formulation does not distinguish between $\hat i\bar{\psi}$ and $i\bar{\psi}$. Thus the combination of $\hat i$ and $i$ in the mass terms reduces to $-1$, yielding real masses. In the context of algebraic spinors, equating $i$ with $\hat i$ is not permitted. Given the Grassmann-odd nature of fermion fields, a mass term such as
\begin{align}
\left\langle \bar{t}t \right\rangle
\end{align}
can be verified to be equivalent to zero, whereas $\hat i\left\langle i\bar{t}t\right\rangle$ is a viable mass term.

The Majorana mass terms (\ref{MIXX}) mix second and third generation neutrinos $\nu_{{\mu}R}$ and $\nu_{{\tau}R}$\footnote{As stated earlier, different flavor projection assignments for relatively subdued interactions are also possible. For example, the following ansatz,
\begin{align}
P^{1} &= P_q\zeta^{0} + P_l\zeta^{+},  \\
P^{2} &= P_q\zeta^{-} + P_l\zeta^{0},  \\
P^{3} &= P_q\zeta^{+} + P_l\zeta^{-},  
\end{align}
can lead to Majorana mass terms mixing $\nu_{eR}$ and $\nu_{\tau R}$ neutrinos, flavor-changing charged interactions between second generation and third generation quarks, and  $h_t$ Higgs boson decaying into tau lepton (among others). Another ansatz,
\begin{align}
P^{1} &= P_q\zeta^{+} + P_l\zeta^{-},  \\
P^{2} &= P_q\zeta^{0} + P_l\zeta^{+},  \\
P^{3} &= P_q\zeta^{-} + P_l\zeta^{0},  
\end{align}
can lead to Majorana mass terms mixing $\nu_{eR}$ and $\nu_{\mu R}$ neutrinos, flavor-changing charged interactions between first generation and third generation quarks, and  $h_t$ Higgs boson decaying into muon (among others).
}, as evidenced in the observation of neutrino oscillations\cite{FUK, AHM, EGU}.  Majorana scale (also called seesaw scale) is much higher than electroweak scale
\begin{align}
&\upsilon_{11}, \upsilon_{23} \gg \upsilon_{t}, \upsilon_{\nu_{\tau}}, \upsilon_{{\tau}}.
\end{align}
Therefore, very small effective masses are generated for neutrinos, known as seesaw mechanism.

Dirac masses within a given condensation cohort ($\upsilon_{t}$, $\upsilon_{\nu_{\tau}}$, or $\upsilon_{{\tau}}$) are proportional to four-fermion coupling constants. As demonstrated in last subsection, the coupling constants are hierarchical. As a result, Dirac masses follow the hierarchy pattern as shown in table~\ref{hierarchy}, where masses in each column are in descending order. Each Higgs boson contributes to masses of four fermions in the same column. The unknown neutrino masses in the table are meant to denote Dirac masses, rather than the significantly smaller seesaw effective masses. One prediction of the current paper is thus the boundaries of neutrino Dirac masses in the table, albeit they have not been determined so far experimentally. Note that the Dirac mass of muon neutrino is expected to be much smaller than the Dirac masses of electron neutrino and tau neutrino. 

\begin{table}[H]
\tbl{Three species of composite electroweak Higgs bosons and corresponding fermion cohorts. Fermion masses are in units of MeV. Error margins of fermion masses are not shown.}
{\begin{tabular}{|lr|lr|lr|}
\hline
\multicolumn{2}{|c|}{Top Quark Higgs} & \multicolumn{2}{c|}{Tau Neutrino Higgs} & \multicolumn{2}{c|}{Tau Lepton Higgs} \\ 
\hline
&&&&& \\
$t$ &173,000 & $\nu_{\tau}$ &? & ${\tau}$ &1,780 \\ 
&&&&& \\
${{\nu_e}}$ &? & ${{c}}$ &1,280 & ${{s}}$ &96 \\ 
&&&&& \\
${b}$ &4,180 & ${\mu}$ &106 & ${{\nu_\mu}}$ &?\\ 
&&&&& \\
${e}$ &0.51 & ${d}$ &4.6 & ${u}$ &2.2 \\
&&&&& \\
\hline
\end{tabular} \label{hierarchy}}
\end{table}

%\vspace{10pt}
Since the coupling constants $g_{t}$, $g_{\nu_{\tau}}$, and $g_{{\tau}}$ are of the same order, the mass discrepancies between top quark, tau neutrino, and tau lepton are driven primarily by the condensation magnitudes $\upsilon_{t}$, $\upsilon_{\nu_{\tau}}$, and $\upsilon_{{\tau}}$. Knowing that top quark is much heavier than tau lepton, we can deduce 
\begin{align}
&\upsilon_{t} \gg \upsilon_{{\tau}}.
\end{align}
Hence top condensation dwarfs tau lepton condensation in terms of contributing to electroweak scale. 

Without knowing Dirac mass of tau neutrino, we can only estimate the relative magnitude of $\upsilon_{\nu_{\tau}}$ based on circumstantial evidences. Given that charm quark mass is much larger than strange quark mass, we hypothesize that\footnote{How do the hierarchies $\upsilon_{11}, \upsilon_{23} \gg \upsilon_{t}, \upsilon_{\nu_{\tau}} \gg \upsilon_{{\tau}}$ (and for that matter, $g_{\nu_{\tau}c} \gg g_{\nu_{\tau}\mu}$) stack up against naturalness principle? One might argue for the hierarchy $\upsilon_{11}, \upsilon_{23} \gg \upsilon_{t}, \upsilon_{\nu_{\tau}}, \upsilon_{{\tau}}$ based on the fact that Majorana masses are not protected by standard model symmetries. However, a reasoning in the same vein would favor a reversed hierarchy by appealing to the fact that Dirac masses are not protected by symmetries $U(1)_{LR} \times U(1)_{B-L}$, which are broken by Majorana masses. We know that electroweak symmetry breaking gives rise to masses of three gauge fields ($W^{\pm}$ and $Z$), while Majorana symmetry breaking contributes to only one gauge field mass ($Z'$). Other than that, the above hierarchies are just assumed. Alternatively, one might assume that $\upsilon_{11}$, $\upsilon_{23}$, and $\upsilon_{t}$ are of the same order, but there is a hierarchy between $G_{11}$/$G_{23}$ and the other four-fermion coupling constants. An interaction such as  $G_{11}\left\langle(\bar{\nu}_{eR} + \bar{t}_{R})\Gamma_2\Gamma_3(\nu_{eR} + {t}_{R})(\bar{\nu}_{eR} + \bar{t}_{R})\Gamma_2\Gamma_3(\nu_{eR} + {t}_{R})\right\rangle$ has $SU(4)$ symmetry for the combination of ${\nu}_{eR}$ and ${t}_{R}$, which is not shared by the right-left-mixing four-fermion interactions. Based on naturalness principle, $g_{\dots}$ can thus be much smaller than $G_{11}$/$G_{23}$.}
\begin{align}
&\upsilon_{\nu_{\tau}} \gg \upsilon_{{\tau}}.
\end{align}
Consequently, tau neutrino condensation might play a substantial role in electroweak scale saturation, as originally envisioned by Martin\cite{NEU1}. Composite Higgs boson $h_{\nu_{\tau}}$ has a sizable branching ratio into charm quark. We anticipate detection of Higgs boson $h_{\nu_{\tau}}$ via the charm quark decay channel, in addition to the 125 GeV Higgs boson\cite{H125A, H125C}, which most likely corresponds to top condensation. Also, in light of the intrinsic connection between muon and Higgs boson $h_{\nu_{\tau}}$, it is worthwhile to investigate tau neutrino condensation's contribution to the muon anomalous magnetic moment\cite{MUON1, MUON2}.

\subsection{Antisymmetric condensation}
Two-fermion condensation might also involve an antisymmetric tensor component\cite{WL2}, such as
\begin{align}
&\Lambda^{-2}\hat i{q}^3_{L}i\bar{t}_{R} \quad\rightarrow\quad \upsilon_{tAT}\gamma_1\gamma_2P_+\zeta^0, \label{LAT}
\end{align}
in addition to the scalar component $\upsilon_{t}P_+\zeta^0$ discussed in last subsection.

The magnitude of this condensation $\upsilon_{tAT}$ could be extremely small compared with the scalar counterpart $\upsilon_{t}$, rendering $\upsilon_{tAT}$-related effects unobservable in laboratories. The miniscule ratio of $\upsilon_{tAT}/\upsilon_{t}$ is in line with naturalness principle, since $\upsilon_{tAT}$ breaks Lorentz (and rotational) symmetry on top of breaking electroweak symmetry. 

The ethereal antisymmetric condensation might manifest itself as 'dark torsion' (or 'dark spin current')\cite{WL2} via interaction with gravitational spin connection. Corrections to torsion and Lorentz violation effects could in turn modify the behavior of gravity on galactic and cosmological scales\cite{MODC}. The antisymmetric condensation (\ref{LAT}) suggests a preferred direction $\gamma_1\gamma_2$ in the universe, which might be reflected as large-scale anisotropies in the cosmic microwave background (CMB)\cite{CMB1, CMB2, CMB3, CMB4}.

\section{Conclusion}
The Higgs sector can be regarded as an effective description of the low energy physics represented by composite boson fields. A challenge facing the composite Higgs model is to account for the vast range of fermion masses which span five orders of magnitude. The current paper is an effort towards explaining the fermion mass hierarchies in the context of composite electroweak Higgs bosons. 

Our approach is based on the framework of six-dimensional Clifford algebra $C\!\ell_{0,6}$ and ternary Clifford algebra element $\zeta$. Standard model fermions can be represented by algebraic spinors of $C\!\ell_{0,6}$,  while ternary $\zeta$-related flavor projection operators dictate allowable flavor-mixing interactions. The first three $C\!\ell_{0,6}$ vectors ($\Gamma_1, \Gamma_2, \Gamma_3$) can be figuratively regarded as cube roots of time dimension associated with trivector $\gamma_0$. The ($\Gamma_1, \Gamma_2, \Gamma_3$)-related trialities include the three colors of quarks and three bivectors of the left-handed weak gauge fields, whereas the three generations of fermions and three composite Higgs bosons are connected with a different sort of triality depicted by ternary Clifford algebra. 

The Lagrangian of the algebraic spinors accommodates $SU(3)_c \times SU(2)_{L} \times U(1)_{R} \times U(1)_{B-L}$ local gauge symmetries. We propose that there are two global chiral symmetries, $U(1)_\alpha$ and $U(1)_\beta$, in addition to the local gauge symmetries. By virtue of dynamical symmetry breaking mechanism, eight-fermion interactions are strong enough to induce four-fermion condensations. These condensations break the global chiral symmetries, while leaving the local gauge symmetries intact. The symmetry-breaking coupling constants of four-fermion interactions are resulted from the four-fermion condensations. Due to the explicit symmetry breaking originated from quantum anomaly and instanton effects, the pseudo-Nambu-Goldstone bosons associated with the four-fermion condensations acquire masses in a similar way as the axion. Since the four-fermion condensations are electroweak singlets, they are more in line with the invisible axion models. It is worthwhile to study further the four-fermion condensations as dark matter candidates and as a possible solution to strong CP problem. 

Two Majorana and three electroweak two-fermion condensations are engendered by four-fermion interactions. Jointly, they break the remaining local gauge symmetries further down to $SU(3)_c \times U(1)_{em}$. Because of the flavor projection operator properties, three or more electroweak condensations are required to generate Dirac masses for all three generations of fermions. We adopt the minimalist approach by allowing for top quark, tau neutrino, and tau lepton condensations only. Each electroweak condensation gives rise to Dirac masses of four different fermions. In accordance with naturalness principle, the global chiral symmetries $U(1)_\alpha$ and $U(1)_\beta$ are instrumental in determining the relative magnitudes of four-fermion coupling constants, and consequently establishing fermion mass hierarchies. One prediction of the current paper is the allowable ranges of different neutrino Dirac masses. For instance, the Dirac mass of muon neutrino is expected to be much smaller than the Dirac masses of electron neutrino and tau neutrino.

Both top quark and tau neutrino condensations play a significant role in electroweak scale saturation. Tau neutrino condensation may also contribute substantially to the muon anomalous magnetic moment. In addition to the 125 GeV Higgs boson observed at the Large Hadron Collider, we anticipate detection of tau neutrino Higgs boson via the charm quark decay channel. On the other hand, a feeble antisymmetric condensation breaks Lorentz symmetry on top of breaking electroweak symmetry. It might be gravitationally relevant and reflected as large-scale CMB anisotropies.

The $C\!\ell_{0,6}$ algebraic spinor potentially allows for $SU(4)$ gauge symmetry, which encompasses $SU(3)_c \times U(1)_{B-L}$. If the $SU(4)$ gauge is taken into consideration as well, the particular flavor projection operator assignment in this paper permits gauge field-driven transformations of first generation quarks into second generation leptons, instead of into first generation leptons as usually assumed for proton decay in grand unified theories. Given the heavy mass of muon, the tree-level amplitude for proton decay into meson and positive muon would therefore be suppressed.

\section*{Acknowledgments}
I am grateful to Gorazd Cvetic, Matej Pavsic, and Greg Trayling for helpful correspondences.

\end{document}